\documentclass[twocolumn,showpacs,preprintnumbers,floatfix]{revtex4}

\usepackage{graphicx}
\usepackage{dcolumn}
\usepackage{bm}

\begin{document}

\title{Charmed and Bottom Baryons from Lattice NRQCD}

\author{Nilmani Mathur$^{a,b}$, Randy Lewis$^{c}$ and R. M. Woloshyn$^{a}$}

\affiliation
{\centerline{$^{a}$TRIUMF, 4004 Wesbrook Mall, Vancouver, British Columbia, Canada V6T 2A3}\\
\centerline{$^{b}$Department of Physics and Astronomy, University of 
Kentucky, Lexington, KY 40506-0055}\\
\centerline{$^{c}$Department of Physics, University of Regina, Regina, Saskatchewan, Canada S4S 0A2}\\}

\begin{abstract}
The mass spectrum of charmed and bottom baryons has been computed on 
anisotropic lattices using quenched lattice nonrelativistic QCD. 
Masses are extracted by using mass splittings which are more accurate 
than masses obtained directly by using the nonrelativistic mass-energy 
relation.  Of particular interest are the mass splittings between spin-1/2
and spin-3/2
heavy baryons, and we find that these color hyperfine effects
are not suppressed in the baryon sector although they are known to be
suppressed in the meson sector.
Results are compared with those obtained in a previous NRQCD calculation 
and with those obtained from a Dirac-Wilson action of the D234 type.
\end{abstract}
\maketitle


\section{INTRODUCTION}
A comprehensive knowledge about the mass spectrum and spin splittings of heavy
baryons is important for our understanding of quantum chromodynamics. 
However, except for singly heavy charmed baryons
and only one singly heavy bottom 
baryon ($\Lambda_{b}$),
most of the heavy baryon masses have not yet been measured
experimentally \cite{pdg}. On the theoretical side there are many results
on heavy baryon masses from different models including,
for example, a number of quark model variations\cite{hv_ex1,hv_ex2,hv_ex3}.

Using lattice QCD,
substantial work has been done in the heavy meson sector. However, so far only 
very few results have been reported for heavy baryons\cite{hv1,hv2,hv3,hv4},
and there is only one work \cite{hv2} where heavy (bottom) quarks are 
treated nonrelativistically.
A further study of charmed and bottom heavy baryons on the lattice using
nonrelativistic QCD (NRQCD) therefore seems worthwhile.

Extraction of the experimentally observed
mass splittings between vector 
and pseudoscalar mesons
remains a challenging problem in lattice QCD;
quenched calculations have so far been unable to extract the observed
mass splittings\cite{quen_res}, and unquenched studies have
not resolved the issue\cite{unquen_res}. Therefore, it is natural to
ask whether lattice results for 
baryon mass splittings also exhibit similar suppression compared to 
experiment.

Empirically, spin splittings in baryons are smaller 
than those in the meson sector. Moreover, in a lattice simulation
the correlators for baryons,
particularly for spin 3/2 states, are noisier than those for mesons, 
and thus, by using lattice QCD it is comparatively difficult to extract a 
reliable mass spectrum for heavy baryons.
In this work we report on the charmed and bottom baryon mass spectrum and mass 
splittings by using a nonrelativistic heavy quark action and an improved 
light quark action on anisotropic lattices.

In section II we summarize different charmed and bottom heavy baryons with 
their relevant quantum numbers and discuss our choice for
interpolating fields.
Section III presents numerical simulation details. 
For heavy quarks, we use the nonrelativistic action from
Ref.~\cite{spwave}, while a tadpole improved gauge action and an improved 
Dirac-Wilson action of the D234 type \cite{alford} are used for light quarks. 
Since these actions were previously detailed elsewhere \cite{hv3,hv4}, we will 
describe them only in an appendix. The calculations are 
done on two different anisotropic lattices with the same gauge configurations 
as were used in Ref.~\cite{hv4} at $\beta = 2.1$ 
and $\beta = 2.3$. 

In section IV we present our results. Masses are calculated using two 
methods; the first uses the standard NRQCD relation between mass and 
energy while the second employs mass splittings to 
calculate masses.
As mass splittings can be estimated more accurately than masses, 
errors in the second method are smaller than those obtained from
the first one.  
The overall systematic uncertainty is estimated by including scale uncertainty,
uncertainty due to the choice of a time window for 
fitting correlation functions, 
error due to extrapolation to the physical light quark masses, uncertainty 
in fixing charm and bottom masses and uncertainty from our determination of the
lattice anisotropy.

Spin splittings are discussed in section V. From our results, along with other 
published results, we conclude that the suppression of mass splittings is 
{\it not present} in the baryon sector in the same way as it is in the
meson sector. 
Over the whole mass range where data are available, quenched 
lattice QCD simulations yield mass differences between spin 3/2 and spin 1/2
baryons which are comparable to or larger than experimental values.

\section{CHARMED and BOTTOM  BARYONS}
Singly and doubly charmed and bottom baryons are summarized in Tables I 
and II respectively. Table II also includes doubly heavy states containing two 
different heavy quarks (charmed and bottom quarks together). Quark content,
as well as the  spin-parity $J^{P}$, the  isospin $I$, and $s_{l}$ 
which identifies the total spin of the light quarks
(also spin-flavor symmetry: $s_{l} = 0$ is symmetric 
while $s_{l} = 1$ is antisymmetric) are shown. Notice that masses
for many 
singly heavy states are not measured yet and there are no data at all on masses 
for doubly heavy states.

\begin{table}
\caption{Summary of singly heavy baryons,
showing valence quark content ($q \equiv u,d$ and $Q \equiv c,b$), 
spin-parity, isospin and mass (in GeV). The quantity $s_{l}$ is the 
total spin of the light quark pair. 
The experimental values are from  Ref.~\protect\cite{pdg}.}

\centerline{}
\begin{ruledtabular}
\begin{tabular}{ccccccc}
Baryons &quark content&$J^{P}$&$I$&$s_{l}$&Mass(c)&Mass(b)\\
\hline
$\Lambda_{Q}$&$udQ$&${1\over2}^{+}$&$0$&$0$&$2.285(1)$&$5.624(9)$\\
$\Xi_{Q}$&$qsQ$&${1\over 2}^{+}$&${1\over 2}$&$0$&$2.468(2)$&\\
$\Sigma_{Q}$&$qqQ$&${1\over2}^{+}$&$1$&$1$&$2.453(1)$&\\
$\Xi^{\prime}_{Q}$&$qsQ$&${1\over2}^{+}$&${1\over 2}$&$1$&$2.575(3)$&\\
$\Omega_{Q}$&$ssQ$&${1\over2}^{+}$&$0$&$1$&$2.704(4)$&\\
$\Sigma^{*}_{Q}$&$qqQ$&${3\over2}^{+}$&$1$&$1$&$2.518(2)$&\\
$\Xi^{*}_{Q}$&$qsQ$&${3\over2}^{+}$&${1\over 2}$&$1$&$2.645(2)$&\\
$\Omega^{*}_{Q}$&$ssQ$&${3\over2}^{+}$&$0$&$1$&&\\
\end{tabular}
\end{ruledtabular}
\end{table}

To project out heavy baryon states we use the same interpolating 
operators as were used in Ref.~\cite{hv4}. For $\Sigma$-like baryons we choose
\begin{equation}
\Sigma \quad : \quad \epsilon ^{abc}[q^{T}_{a}C\gamma _{5}Q_{b}]q_{c}
\end{equation}
where $q$ is a light quark field and $Q$ is a heavy quark field.
Here $a,b,c$ are color indices whereas Dirac indices have been suppressed.
For $\Sigma_{Q}$, $q$ is $u$ or $d$ and for $\Omega_{Q}$, $q$ is $s$. 
For doubly heavy $\Sigma$-like baryons with equal heavy masses, we interchange
the role of light and heavy fields, {\it i.e.}, to get $\Xi_{QQ}$, we change 
$q \rightarrow Q$ and $Q \rightarrow u$ or $d$. Similarly, for $\Omega_{QQ}$, 
the change is $q \rightarrow Q, Q \rightarrow s$.

The $\Xi^{\prime}_{Q}$ is $\Sigma$-like but it contains 
two different light flavors so it is considered separately as 
\begin{eqnarray}
\Xi^{\prime}&:& \frac{1}{\sqrt{2}}\left\{ \epsilon ^{abc}[q_{a}'^{T}C\gamma _{5}Q_{b}]q_{c}
+\epsilon ^{abc}[q^{T}_{a}C\gamma _{5}Q_{b}]q_{c}'\right\},
\end{eqnarray}
with $q = u$ or $d$ and $q^{\prime} = s$.

The $\Lambda$-like baryons involve three distinct flavors. A simple choice is 
the $\it heavy \, lambda$ :
\begin{equation}
\Lambda  \quad : \quad \epsilon ^{abc}[q^{T}_{a}C\gamma _{5}q_{b}']Q_{c}.
\end{equation}
where for $\Lambda_{Q}$, $q = u$, $q^{\prime} = d$, and for $\Xi_{Q}$,
$q = u$, $q^{\prime}$ = s. A more symmetrical choice would be the 
{\it octet lambda}  
\begin{eqnarray}
\Lambda_{o} &:&  \frac{1}{\sqrt{6}}\epsilon ^{abc}{\biggl\{} 2[q^{T}_{a}C\gamma _{5}q_{b}']Q_{c}
+ [q^{T}_{a}C\gamma _{5}Q_{b}]q_{c}'\nonumber\\
&&\hspace{1.3in} - [q_{a}'^{T} C\gamma _{5}Q_{b}]q_{c}{\biggr\}},
\end{eqnarray}
with the same flavor assignment as for the {\it heavy lambda}. One can use
either of these $\Lambda$ states as they give consistent results\cite{hv4}. 
We choose the octet-lambda $(\Lambda_{o})$ for this work.
For spin 3/2 baryons we choose the following interpolating field:
\begin{equation}\label{Sigma*}
\Sigma^{*} \quad : \quad \epsilon ^{abc}[q^{T}_{a}C\gamma _{\mu }q_{b}']Q_{c},
\end{equation}
where for $\Sigma^{*}_{Q}$, $q = q^{\prime}$ is $u$ or $d$ and for 
$\Omega^{*}_{Q}$, $q = q^{\prime}$ is $s$. To get $\Xi^{*}_{Q}$, one needs to 
consider $q = u$ or $d$ and $q^{\prime} = s$. Similarly, to get 
the doubly heavy states with equal heavy masses one needs to interchange 
the role of light and heavy fields. For example, to get $\Sigma^{*}_{QQ}$, 
one needs to change $q, q^{\prime} \rightarrow Q$ 
and $Q \rightarrow u$ or $d$, whereas $\Omega^{*}_{Q}$ requires $q, 
q^{\prime} \rightarrow Q$ and $Q \rightarrow s$.
\begin{table}
\caption{Summary of doubly heavy baryons,
showing valence quark content ($q \equiv u,d$ and $Q \equiv c,b$),
spin-parity, isospin and $S_{QQ}$, the total spin of the heavy quark 
pair.}
\centerline{}
\begin{ruledtabular}
\begin{tabular}{cccccc}
&Baryons&quark content&$J^{P}$&$I$&$S_{QQ}$\\
\hline
&$\Xi_{QQ}$&$qQQ$&${1\over2}^{+}$&${1\over 2}$&$1$\\
&$\Omega_{QQ}$&$sQQ$&${1\over2}^{+}$&$0$&$1$\\
&$\Xi^{*}_{QQ}$&$qQQ$&${3\over2}^{+}$&${1\over 2}$&$1$\\
&$\Omega^{*}_{QQ}$&$sQQ$&${3\over2}^{+}$&$0$&$1$\\
\hline
&$\Xi_{bc}$&$qbc$&${1\over2}^{+}$&${1\over 2}$&$0$\\
&$\Omega_{bc}$&$sbc$&${1\over2}^{+}$&$0$&$0$\\
&$\Xi^{\prime}_{bc}$&$qbc$&${1\over2}^{+}$&${1\over 2}$&$1$\\
&$\Omega^{\prime}_{bc}$&$sbc$&${1\over2}^{+}$&$0$&$1$\\
&$\Xi^{*}_{bc}$&$qbc$&${3\over2}^{+}$&${1\over 2}$&$1$\\
&$\Omega^{*}_{bc}$&$sbc$&${3\over2}^{+}$&$0$&$1$\\
\end{tabular}
\end{ruledtabular}
\end{table}

The operator in Eq.~(\ref{Sigma*}) has both spin 1/2 and spin 3/2 states.
At zero momentum the corresponding correlation function can be written as
\cite{benmer}:
\begin{eqnarray}
C_{ij}(t)=(\delta _{ij}-\frac{1}{3}\gamma _{i}\gamma _{j})C_{3/2}(t)+
\frac{1}{3}\gamma _{i}\gamma _{j}C_{1/2}(t),
\end{eqnarray}
where $i, j$'s are spatial Lorentz indices and
$C_{3/2 (1/2)}$ are the spin projections for spin 3/2 (1/2) states. 
By choosing different Lorentz components the spin 3/2 part,
 $C_{3/2}(t)$, is extracted and used to calculate the mass of the spin 3/2
baryons.

Operators for baryons with two unlike heavy flavors may be constructed 
from the above interpolating operators by interchanging the role of 
heavy and light fields. For example, $\Xi^{*}_{QQ^{\prime}}$ 
and $\Omega^{*}_{QQ^{\prime}}$ can be obtained 
from Eq.~(5) by letting $q$, $q^{\prime} \rightarrow Q, Q^{\prime}$ 
and $Q \rightarrow q$ with $q = u$ or $d$ and $q = s$ respectively.
For $\Xi^{\prime}_{QQ^{\prime}}$ and $\Omega^{\prime}_{QQ^{\prime}}$
we use the symmetrical form again, as given by Eq.~(3), 
making the same replacements 
{\it i.e.}, $q$, $q^{\prime} \rightarrow Q, Q^{\prime}$ 
and $Q \rightarrow q$ with $q = u$ or $d$ and $q = s$ respectively. Finally,
$\Xi_{QQ^{\prime}}$ and $\Omega_{QQ^{\prime}}$ are the doubly heavy analogs of $\Lambda$ and they can be obtained from Eqs.~(3) and (4) as previously.

\begin{figure*}[htb!]
\vspace*{-1.4in}
\hspace*{-0.4in}
\includegraphics[height=16.0cm]{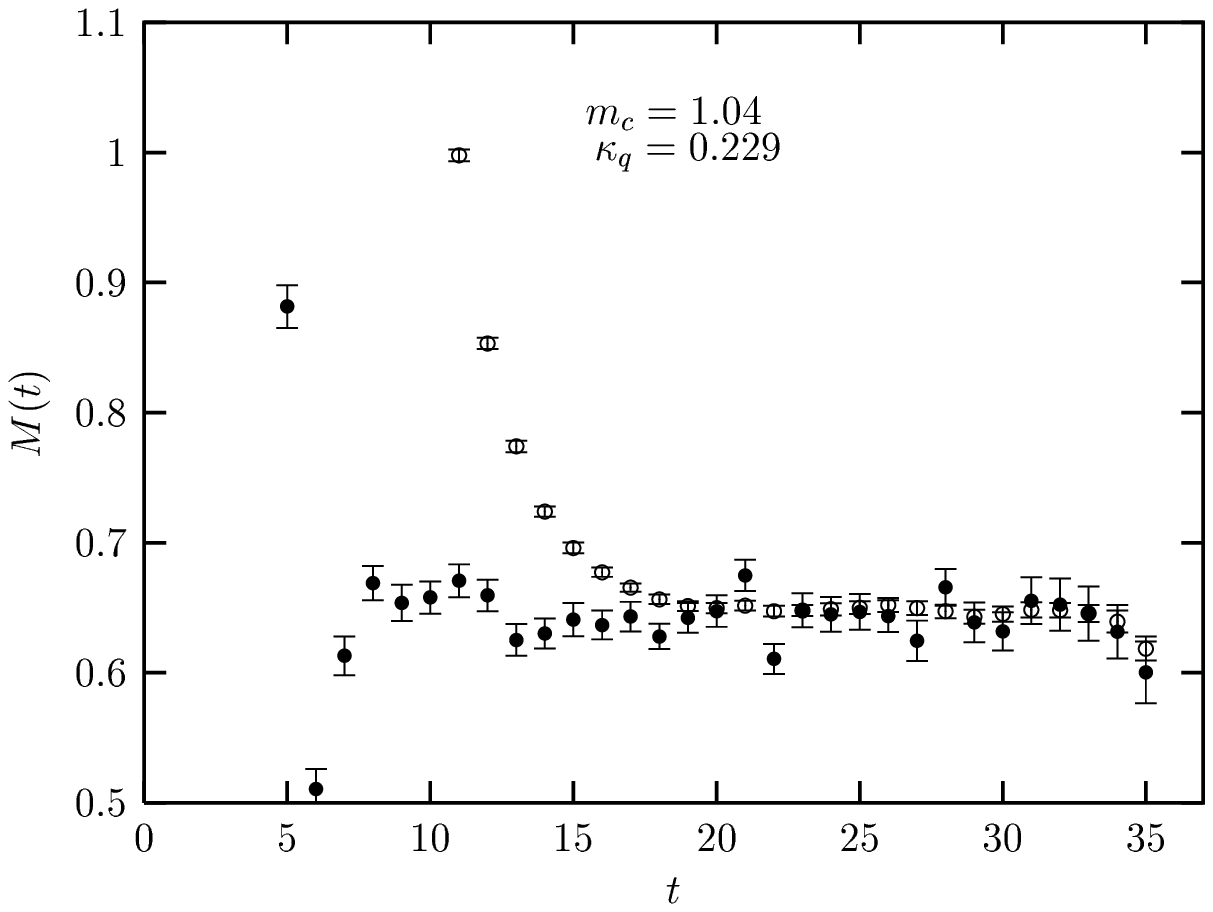}
\hspace*{-1.6in}
\includegraphics[height=16.0cm]{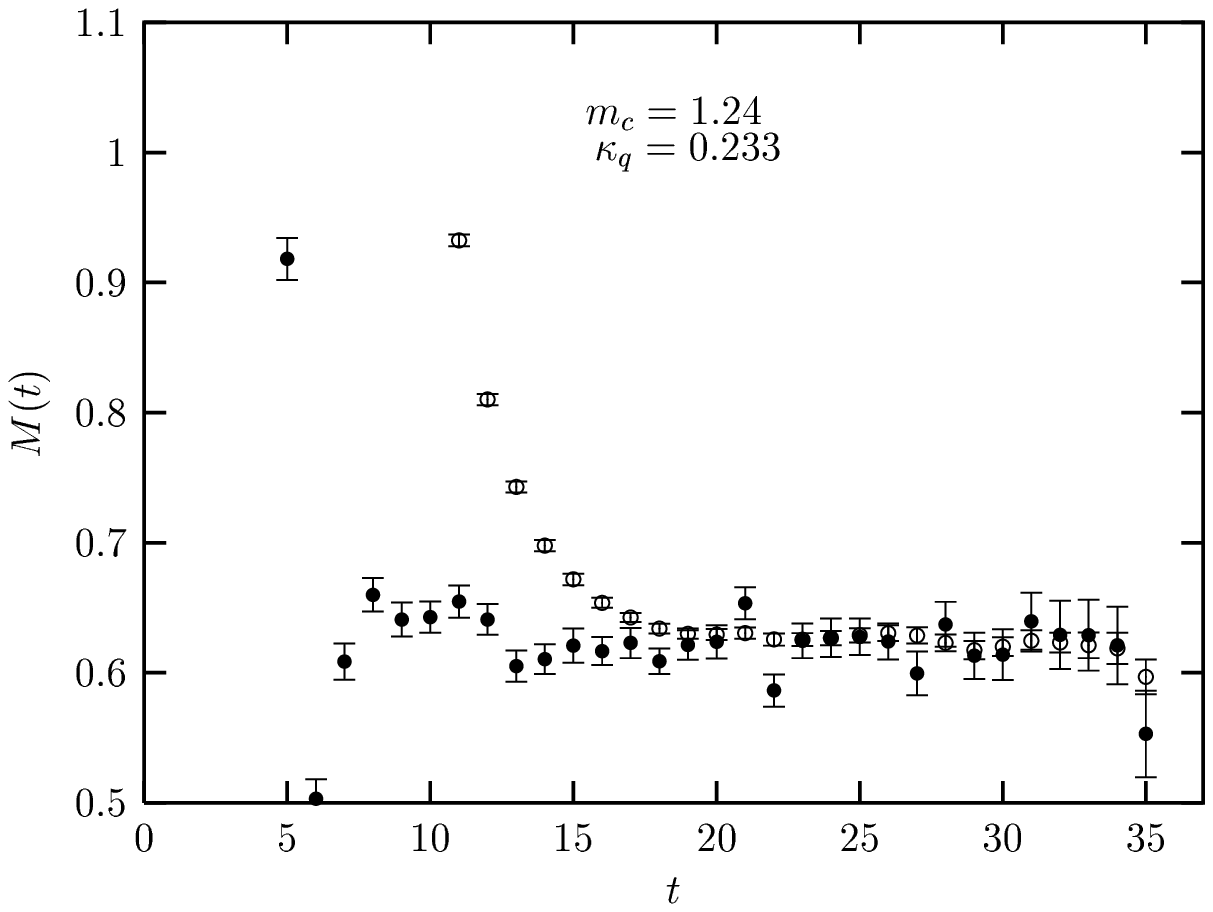}
\\
\vspace*{-4.3in}
\hspace*{-0.4in}
\includegraphics[height=16.0cm]{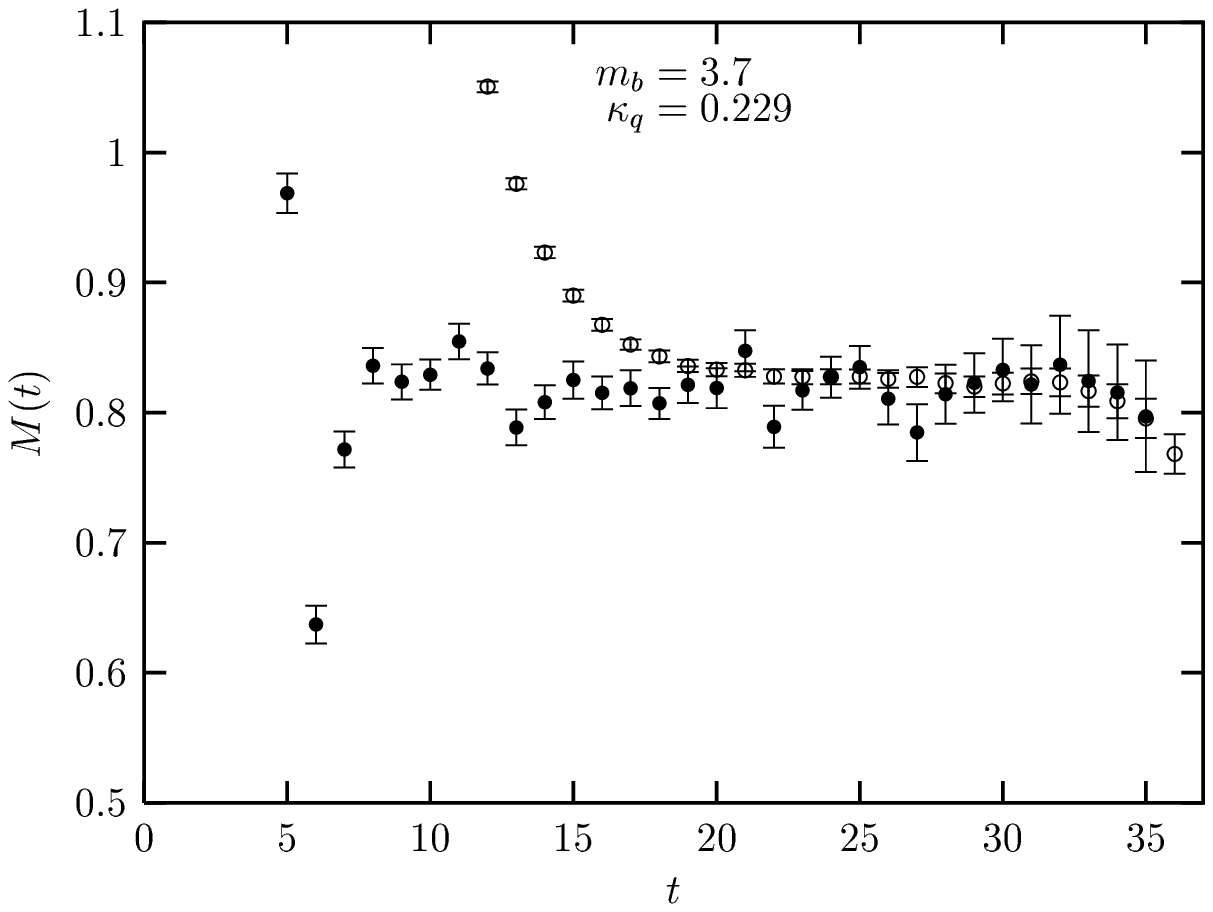}
\hspace*{-1.6in}
\includegraphics[height=16.0cm]{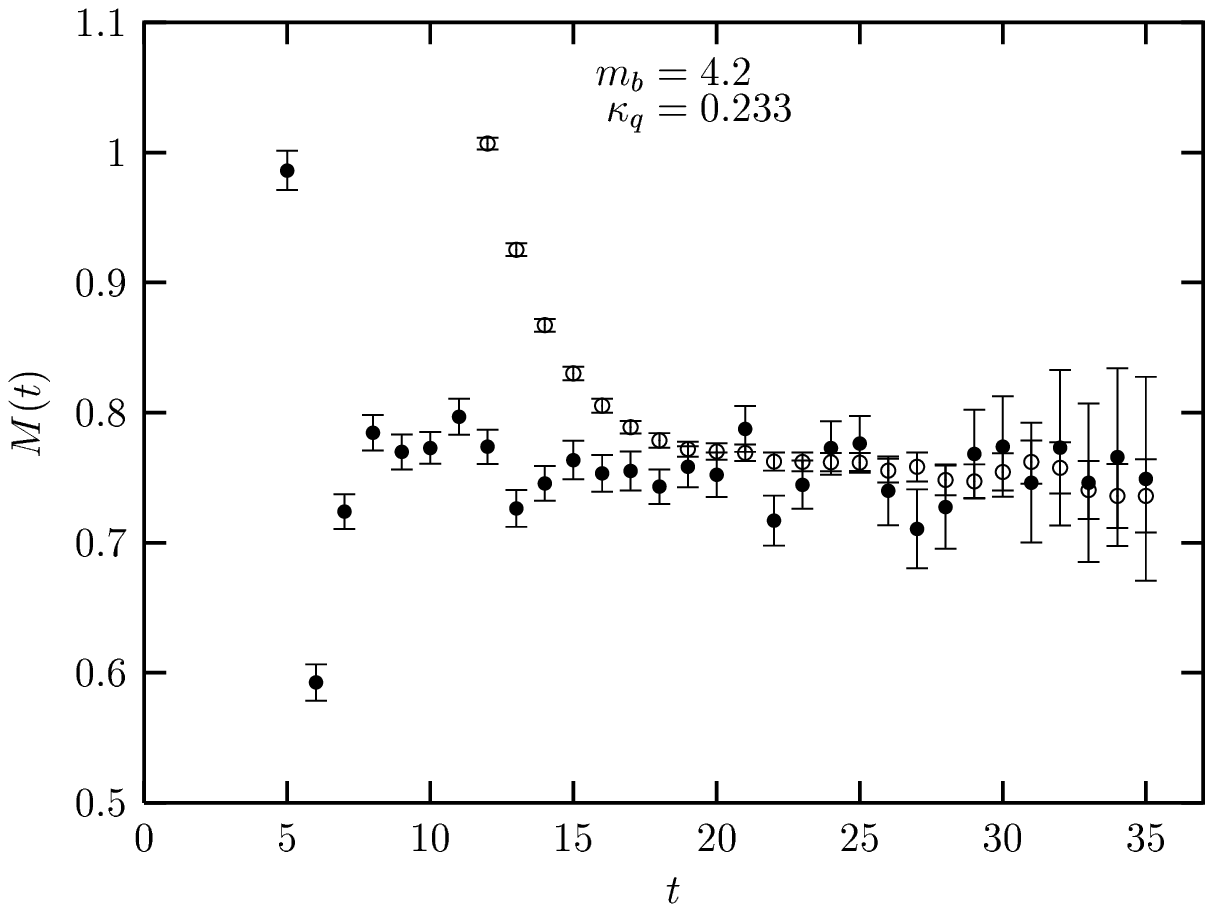}
\vspace*{-3.1in}
\caption{Effective Mass $M(t)$ versus $t$ for singly heavy $\Sigma$-like
 baryons for different combinations of light and heavy
 quark mass (denoted by hopping parameter $\kappa$ and bare mass $m$, 
 respectively).
 Open symbols are for calculations with a correlation
 function with local source and sink, filled symbols are for local
 source and smeared sink.}
\end{figure*}

\section{NUMERICAL SIMULATION}
\subsection{Actions}
The gauge action as well as the heavy quark NRQCD action used for this work
are described in 
detail in Ref.~\cite{spwave}. The gauge action is tadpole improved and the 
leading classical error is quartic in lattice spacing. The Hamiltonian 
corresponding to the NRQCD action
is complete to ${\mathcal{O}}(1/M^3)$ in the classical continuum limit. 
For light quarks we use a Dirac-Wilson action of the D234 type \cite{alford} 
which has been used previously and detailed in Refs.~\cite{hv3,hv4}. Its 
leading classical errors are cubic in lattice spacing. 
All these actions are summarized in the appendix. 

\begin{figure*}[htb!]
\vspace*{-1.4in}
\hspace*{-0.4in}
\includegraphics[height=16.0cm]{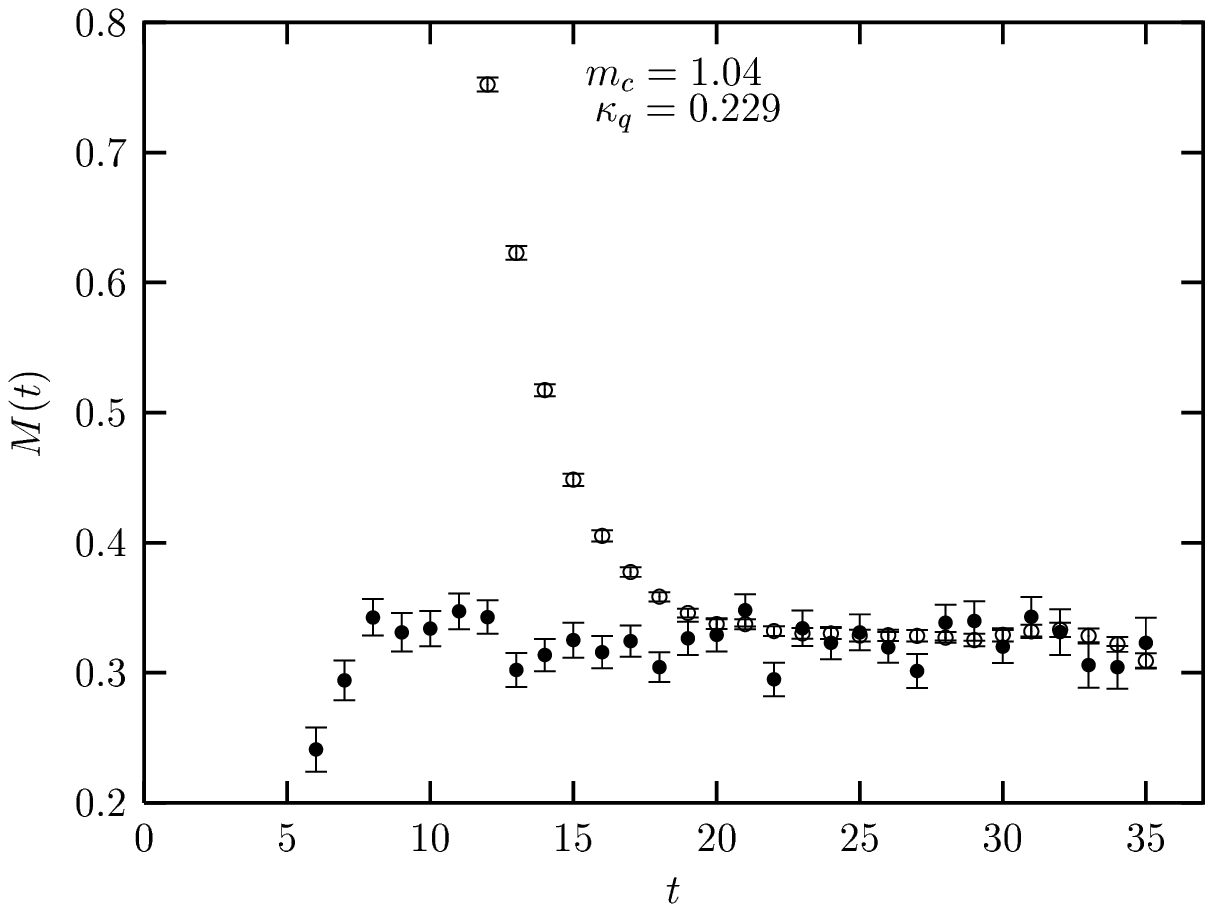}
\hspace*{-1.6in}
\includegraphics[height=16.0cm]{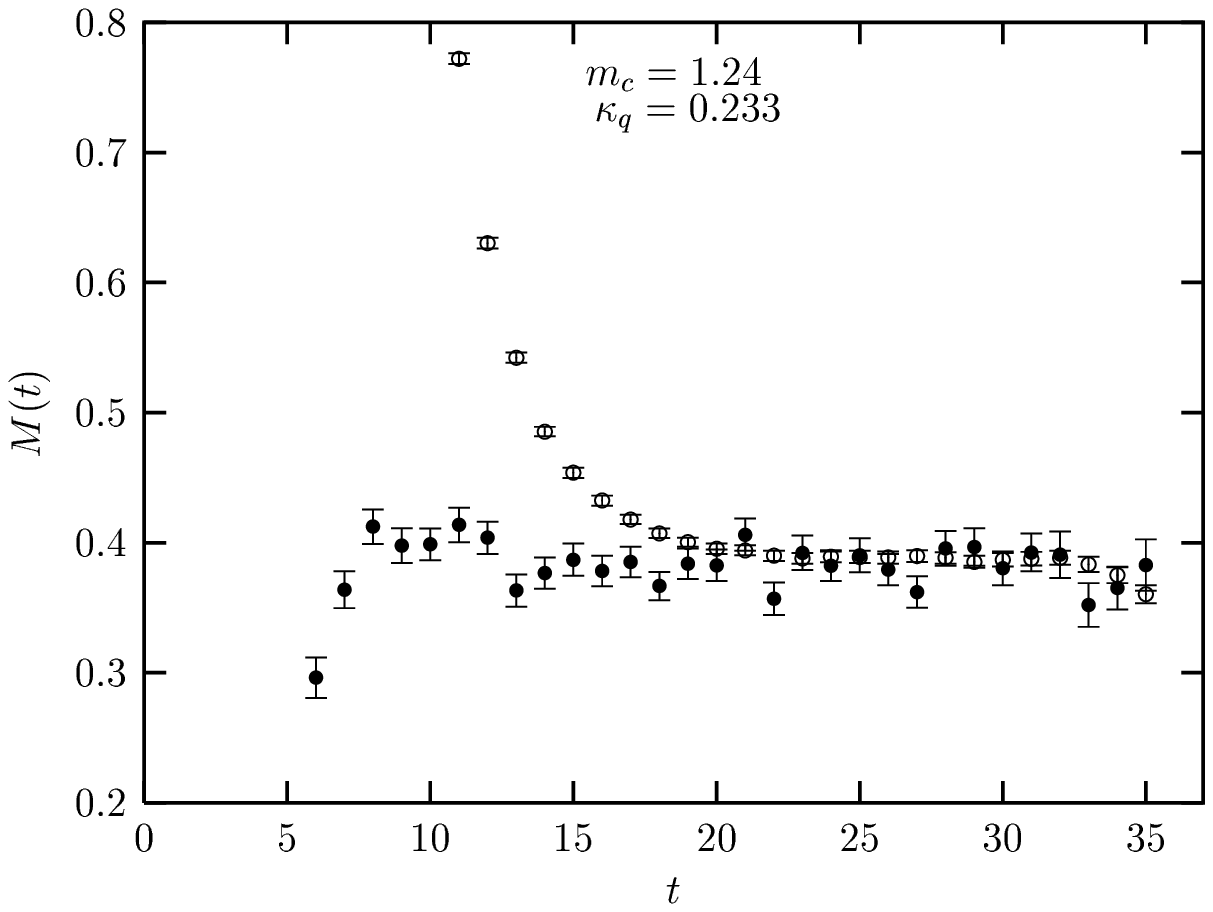}
\\
\vspace*{-4.3in}
\hspace*{-0.4in}
\includegraphics[height=16.0cm]{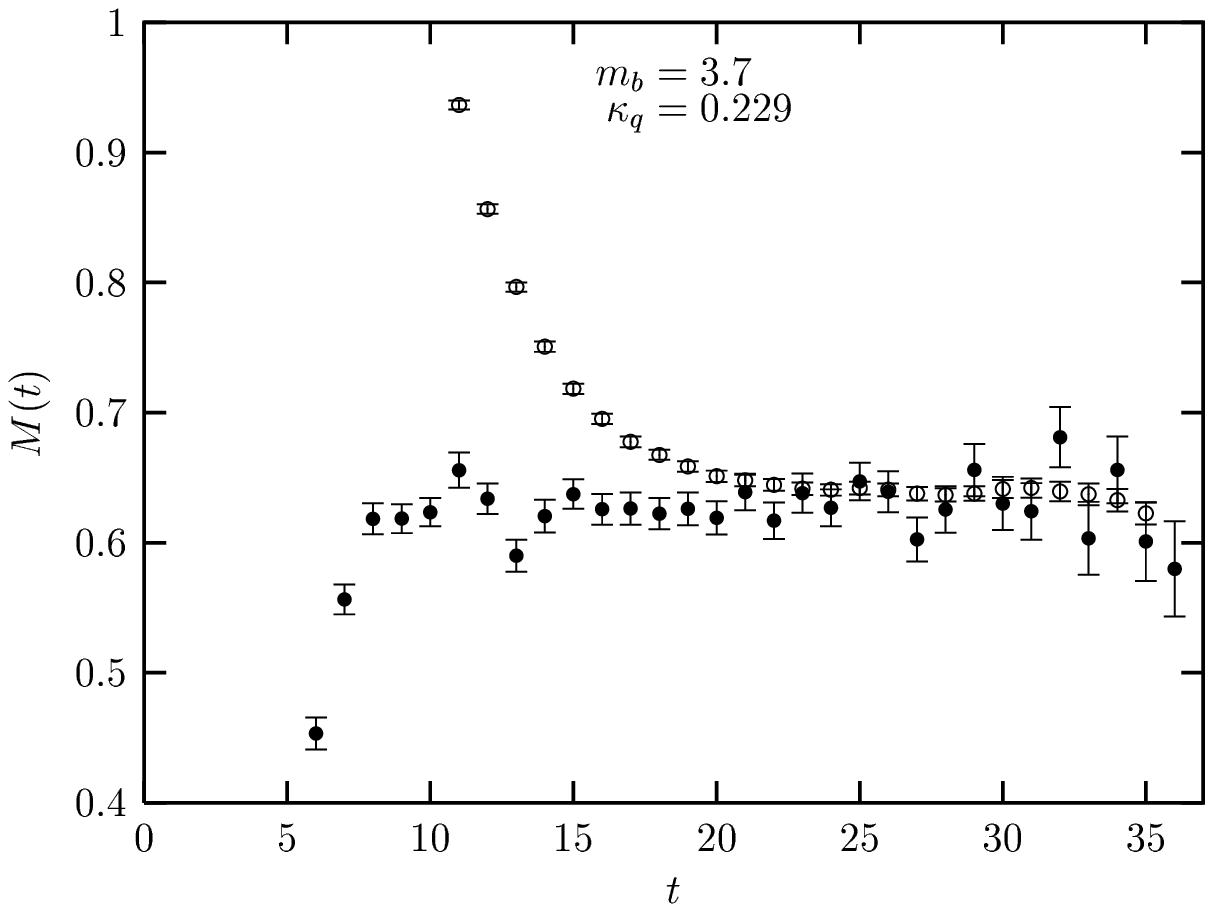}
\hspace*{-1.6in}
\includegraphics[height=16.0cm]{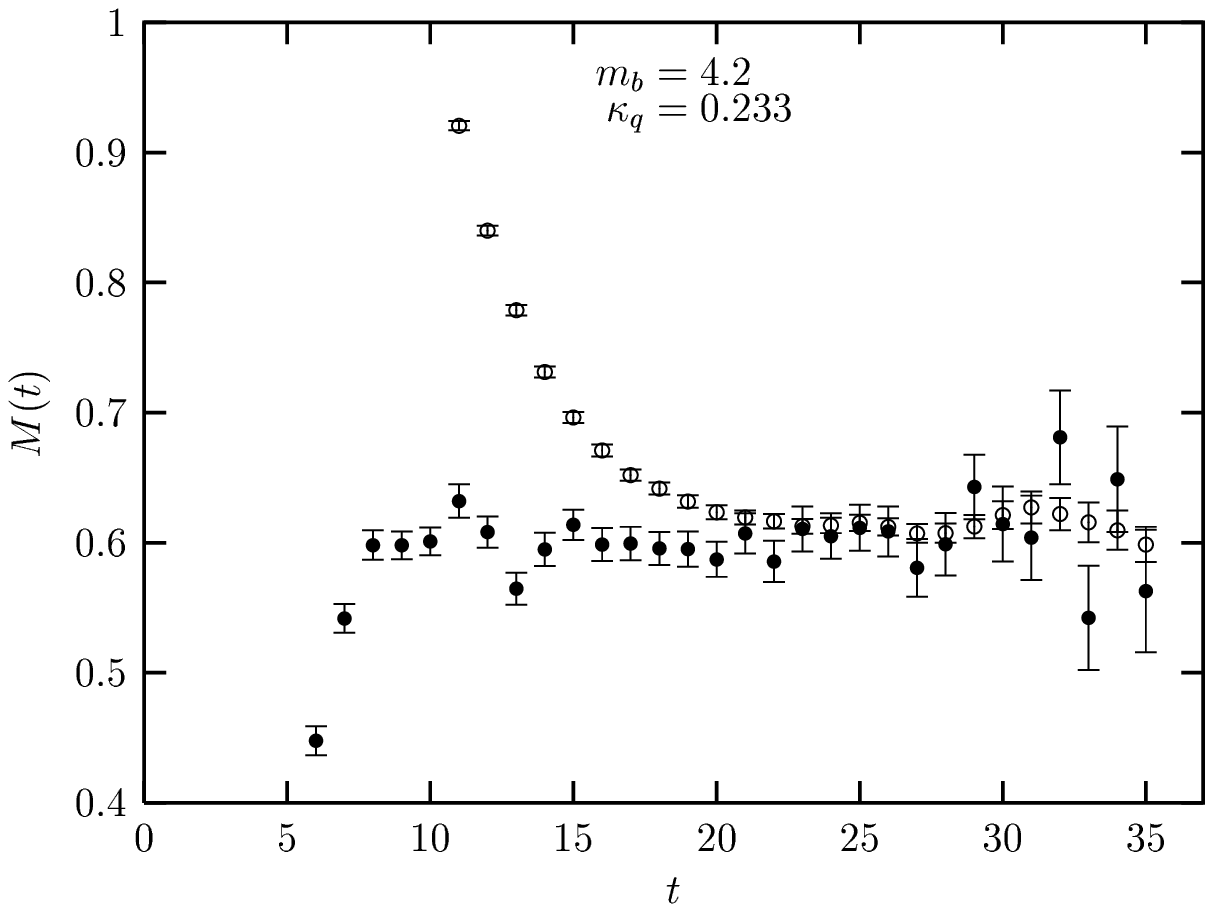}
\vspace*{-3.1in}
\caption{Effective Mass $M(t)$ versus $t$ for doubly heavy $\Sigma$-like 
 baryons for different combinations of light and heavy
 quark mass (denoted by hopping parameter $\kappa$ and bare mass $m$, 
 respectively).
 Open symbols are for calculations with a correlation
 function with local source and sink, filled symbols are for local
 source and smeared sink.}
\end{figure*}

\subsection{Simulation Details}

This work is done with two sets of quenched gauge configurations
(at $\beta=2.1$
and $2.3$) on anisotropic lattices with a bare aspect ratio $a_{s}/a_{t}
= 2$,
where the spatial lattice spacing varies from about 0.22 to 0.15 fm.
 
The renormalized anisotropy is obtained from
\begin{equation}\label{xicalc}
\xi = \frac{a_s}{a_t} = \frac{a_sV(r_2) - a_sV(r_1)}{a_tV(r_2) -
a_tV(r_1)},
\end{equation}
where $V(r)$ is the potential between a static quark-antiquark pair
with separation $r$, and is extracted from an exponential fit to a
sequence of Wilson loops.
In the numerator of Eq.~(\ref{xicalc})
the sequence of Wilson loops extends in a coarsely-spaced direction,
and in the denominator the sequence extends in the finely-spaced
direction.  The separation
$r$ may be along a lattice axis or off-axis, and various possibilities
were included in the calculation.  However,
the separation $r$ never includes the finely-spaced
direction, so that $V(r)$ itself is independent of $a_s/a_t$.
It is convenient to avoid using the largest values of $r$, where
the exponential fit becomes noisier and the periodicity of the lattice
can affect a determination of the anisotropy.  Our results are
\begin{eqnarray}
\xi = \frac{a_s}{a_t} = \left\{ \begin{array}{l}
       1.96(2), {\rm ~for~} \beta=2.1 \\
       1.99(3), {\rm ~for~} \beta=2.3\,.
       \end{array}\right .
\end{eqnarray}

\begin{figure*}[htb!]
\vspace*{-1.4in}
\hspace*{-0.4in}
\includegraphics[height=16.0cm]{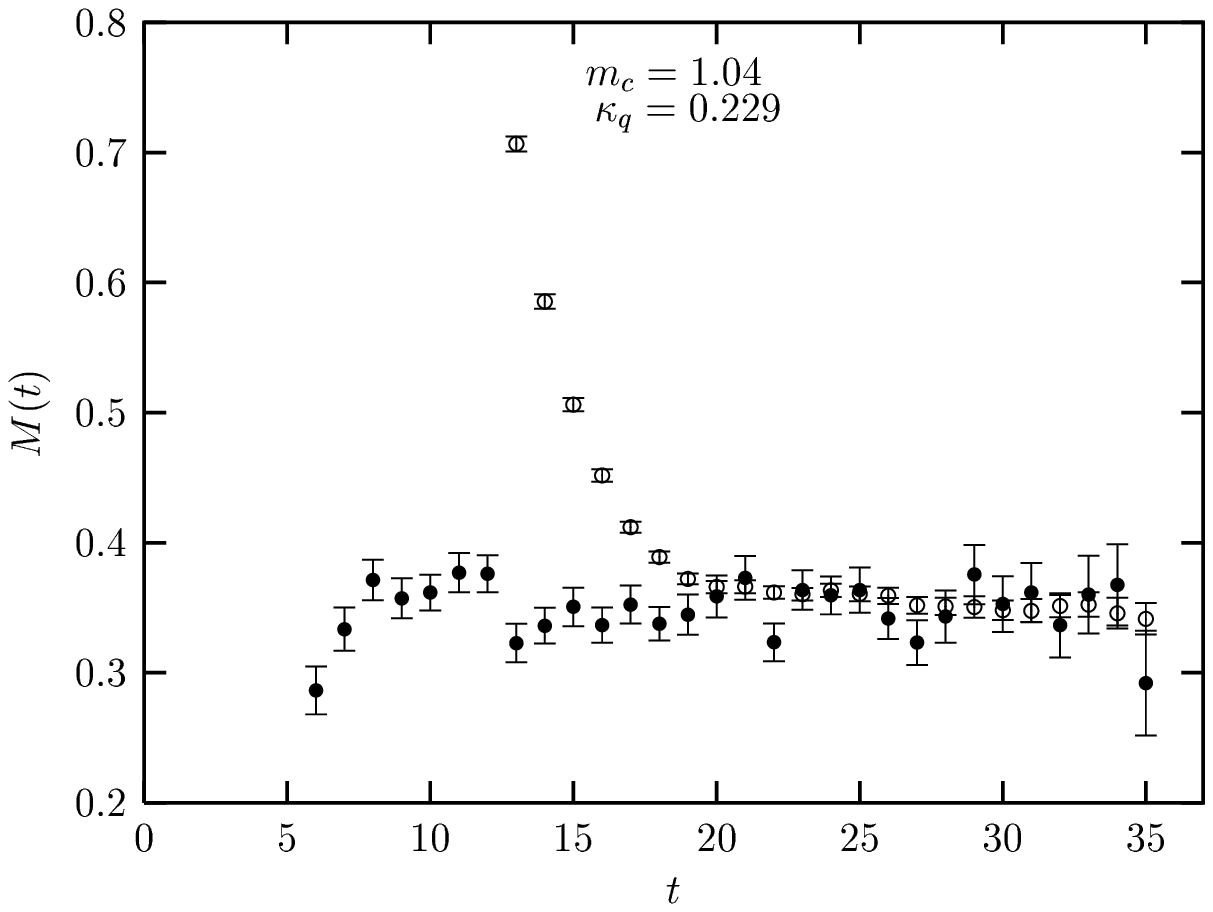}
\hspace*{-1.6in}
\includegraphics[height=16.0cm]{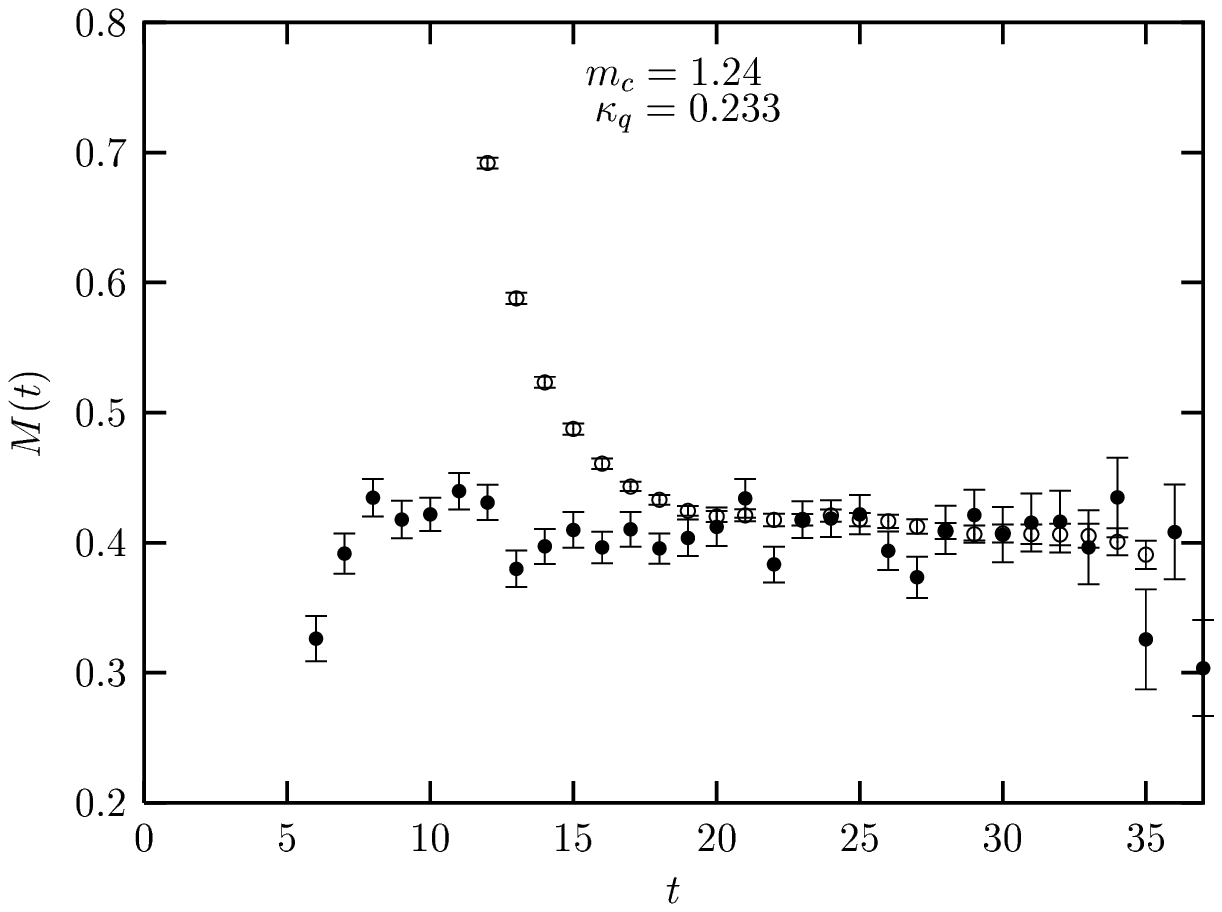}
\\
\vspace*{-4.3in}
\hspace*{-0.4in}
\includegraphics[height=16.0cm]{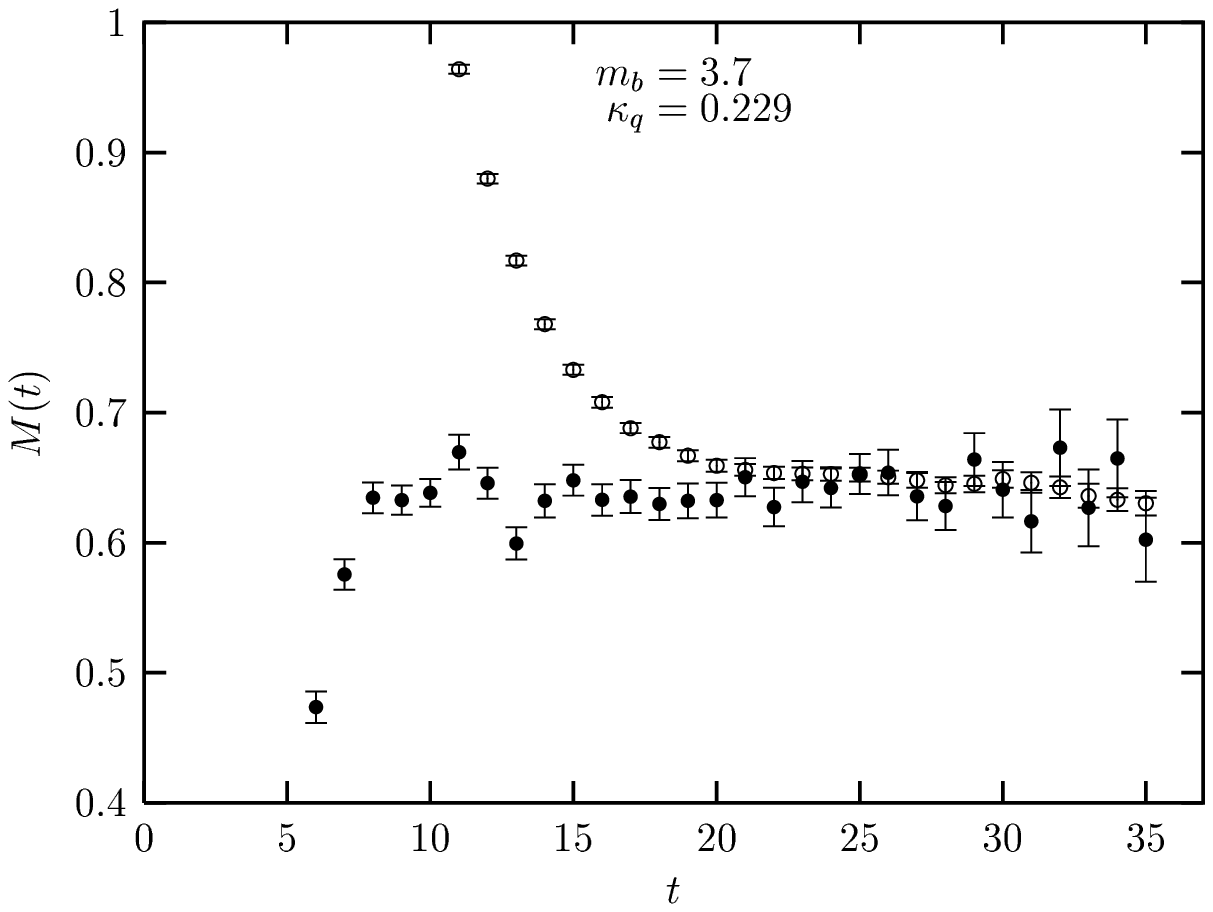}
\hspace*{-1.6in}
\includegraphics[height=16.0cm]{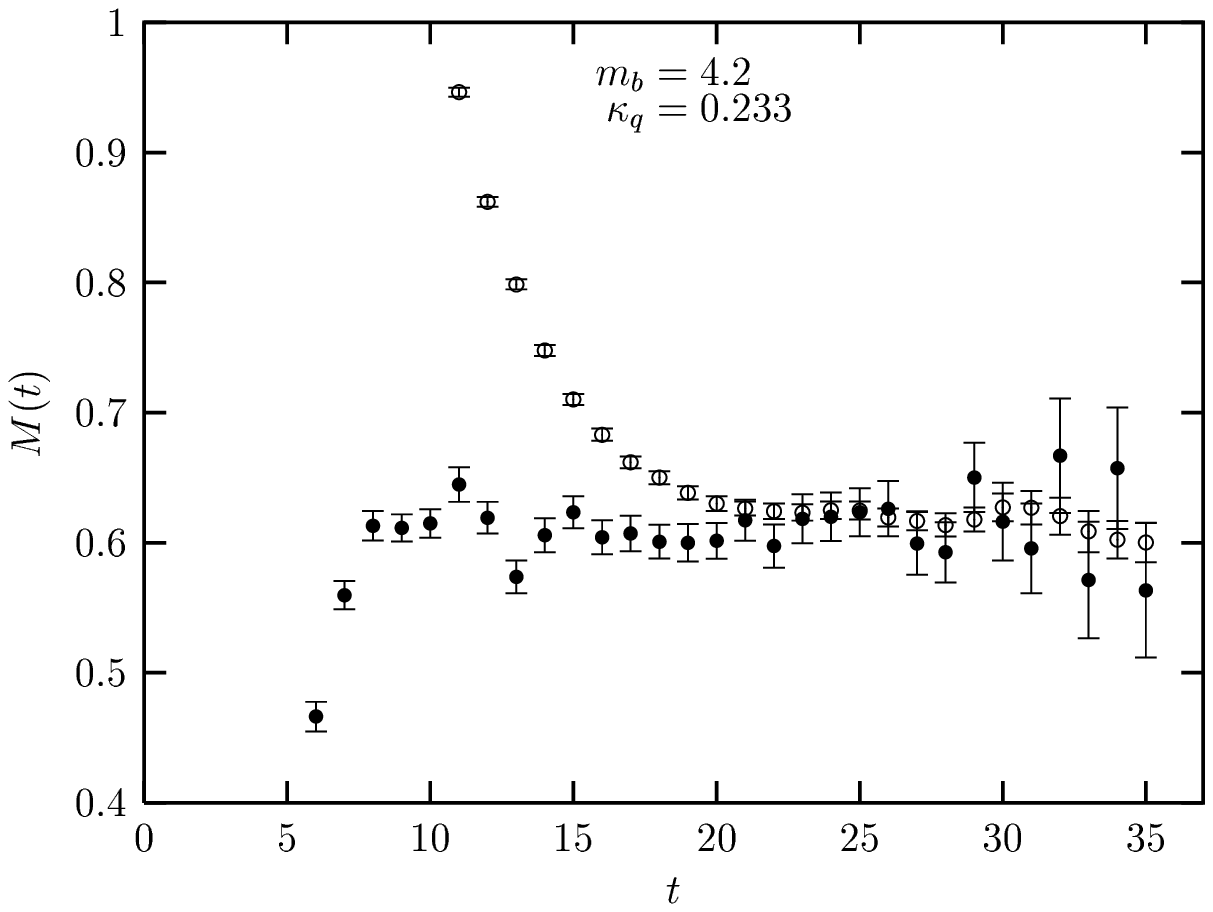}
\vspace*{-3.1in}
\caption{Effective Mass $M(t)$ versus $t$ for doubly heavy $\Sigma^{*}$-like
 baryons for different combinations of light and heavy
 quark mass (denoted by hopping parameter $\kappa$ and bare mass $m$, 
 respectively).
 Open symbols are for calculations with a correlation
 function with local source and sink, filled symbols are for local
 source and smeared sink.}
\end{figure*}

We used fixed time boundaries to construct quark propagators,
and gauge fields were generated using a pseudo-heat-bath Monte Carlo algorithm 
with 400 ($\beta = 2.1$) to 800 ($\beta = 2.3$) sweeps 
between saved configurations.
For $\beta=2.1$, we use 720 configurations and for $\beta = 2.3$ the
number of 
configurations is 442. Two sets of bare masses are used for each heavy quark
while four sets of hopping parameters are used for the light one. Bare masses
for heavy quarks are chosen to surround the physical value so that
an interpolation can be used. For example, at 
$\beta = 2.1$,
the charm mass is in between 1.2 and 1.5 and the bottom mass is 
in between 5.0 and 6.0. The charm mass is fixed by setting the $\eta_c$ mass
to its experimental value, 
whereas the $B^{0}$ mass is used to fix the bottom mass. 
The hopping parameter
corresponding to the strange quark is fixed from the $D_s$ meson mass.
The temporal lattice spacing and correspondingly the scale is fixed by 
setting the $\rho$-meson mass to its experimental value. 
Summaries of lattice parameters as well as hopping parameters 
for heavy and light  fields are given in Tables III and IV respectively. 

\begin{table}
\caption{Summary of lattice parameters. The quantity 
$a_{t}^{-1}$ is the inverse of the temporal lattice
spacing while $u_{s}$ and $u_{t}$ are the tadpole improvement
factors for spatial and temporal links respectively.}
\centerline{}
\begin{ruledtabular}
\begin{tabular}{cccccc}
\( \beta  \)&
size&
configurations&
\( a_{t}^{-1}(GeV) \)&
\( u_{s} \)&
\( u_{t} \)\\
\hline 
2.1&
\( 12^{3}\times 32 \)&
720&
1.803(42)&
0.7858&
0.9472\\
2.3&
\( 14^{3}\times 38 \)&
442&
2.210(72)&
0.8040&
0.9525\\
\end{tabular}
\end{ruledtabular}
\end{table}

Correlation functions are calculated using interpolating operators 
in local form at both source and sink. In addition to that we use
a gauge invariant smearing for quark propagators 
at the sink using
the smearing function from Eq.~(13) of 
Ref.~\cite{smear}. These local and sink-smeared correlators are 
fitted simultaneously to obtain hadron masses. 
The required correlations 
among different quantities are taken into account 
by covariant matrices obtained from singular value decomposition, and the
statistical error is estimated from bootstrapping the fitting procedure.
As in Ref.~\cite{hv4}, local correlators are fitted with two exponential 
functions ($A \exp{(-m_{1}x)} + B \exp{(-m_{2}x)}$), while the sink-smeared 
correlation function is fitted with a single exponential ($C\exp{(-m_{1}x)}$). 
The mass parameter for the sink-smeared fit is constrained to be the same as 
the lowest mass of the fit to the local correlator. 
The time window for the fit is chosen in a way such that 
the ending time is large and the fit is stable under variation of both starting 
and ending time by a few time steps.
\begin{table}
\caption{Hopping parameters and bare masses. Four $\kappa$ values were
used in simulations at each $\beta$.  $\kappa_{s}$ is
the hopping parameter for the strange quark, and
$c$ and $b$ are the charmed and bottom bare masses respectively.}
\centerline{}
\begin{ruledtabular}
\begin{tabular}{ccccc}
$\beta$ &$\kappa$&$\kappa_{s}(\phi)$ &\multicolumn{2}{c}{bare mass}\\
&&&$c$&$b$\\
\hline
2.1&0.229,0.233,0.237,0.240& 0.2338&1.2,1.5&5.0,6.0\\
2.3&0.229,0.233,0.237,0.240& 0.2371&1.04,1.24&3.7,4.2\\
\end{tabular}
\end{ruledtabular}
\end{table}
Light quark extrapolation is done by extrapolating the hadron masses extracted 
at four light quark masses with the form
$c_{0} + c_{2}m_\pi^{2} + c_{3}m_{\pi}^{3}$, where $m_{\pi}$ is the pion 
mass. In most of the cases the cubic ($m_{\pi}^{3}$) contributions are small 
and they are included only to get systematic errors.

Figs.~1-3 show some representative examples of our simulation results. 
We plot the effective mass for different heavy baryons versus time $t$, 
where the effective mass is defined to be $M(t) = ln(g(t)/g(t+1))$ with
$g(t)$ being the zero-momentum time correlation function of baryon fields.
Open symbols in these figures are for calculations with a correlation
function with local source and sink, filled symbols are for local
source and smeared sink. There is good agreement between local and smeared
results at large times. 

It should be noted that the actual fits to determine 
the masses are performed directly with the
correlation functions, and not on the effective masses plotted in 
Figs.~1-3,
but the plots provide an indication of the quality of our data.
Although the sink-smeared results appear to be somewhat noisy they are
quite helpful in constraining the two-exponential fit of the 
local correlation function.

\subsection{Mass Extraction}
The kinetic mass of a nonrelativistic state can be extracted from the usual
NRQCD relation \cite{spwave}
\begin{eqnarray}
M_{kin} &=& {2\pi^{2}\over {N_{s}^{2}\xi^{2}a_{t}[a_{t}(E_{p}-E_{0})]}},
\end{eqnarray}
which is derived from $E = {p^{2}\over 2M_{kin}}$. Here $N = La_{s}$, with
$L$ being the lattice size and $a_{s}$ the spatial lattice spacing. 
$\xi = a_{t}/a_{s}$ is the anisotropy whereas $E_{0}$ and $E_{p}$ are
simulation energies corresponding to the ground state and the state with
momentum $p = {2\pi\over La_{s}}$, respectively.
Mass differences  between two states ($H_{1}$ and $H_{2}$) with the same heavy
quark can be obtained by taking the difference of their zero momentum
simulation energies:
\begin{eqnarray}
M_{H_{1}} - M_{H_{2}}&=&E^{1}_{sim}(0) - E^{2}_{sim}(0),
\end{eqnarray}
which follows from the lattice NRQCD expression for the hadron
mass
\begin{eqnarray}
M_{H}&=&E_{sim}(0) + Z M_{Q} - E_{shift},
\end{eqnarray}
where $E_{sim}$ is the simulation energy at zero momentum and the 
last two terms represent the renormalized heavy quark mass.
The bare quark mass $M_{Q}$ has both a multiplicative ($Z$) and 
additive ($E_{shift}$) renormalization\cite{davthack} which should be 
independent of hadronic state.
A more precise result is obtained for heavy hadron masses with
the heaviest light quark ($\kappa = 0.229$) rather than with a lighter
light quark ($\kappa = 0.233$ and higher). 
\begin{table}[htb!]
\caption{Results for meson and baryon masses and mass splittings (in MeV)
calculated using the NRQCD action for charm ($c$) and bottom ($b$) quarks.
The first error is statistical while the second error
comprises systematic errors due to scale, time window and anisotropy. Rows are
separated 
into mesons, singly heavy charmed baryons, doubly heavy charmed baryons,
singly heavy bottom baryons and doubly heavy bottom baryons, respectively.}
\centerline{}
\begin{ruledtabular}
\begin{tabular}{cll}
&$\beta = 2.1$&$\beta = 2.3$\\
\hline
$J/\Psi - \eta_{c}$ & $70(2)(^{5}_{4})$ 
& $76(3)(^{7}_{5}) $\\
$D$ & $1842(28)(^{33}_{31})$
& $1850(35)(^{28}_{24})$ \\
$D_{s}$ & $1980(23)(^{26}_{23})$ 
&$1958(33)(^{23}_{21})$\\
$D^{*} - D$ &$98(6)(^{5}_{3})$
&$101(6)(^{6}_{5})$\\
$D^{*}_{s} - D_{s}$ &$94(4)(^{4}_{3})$
&$96(4)(^{4}_{5})$\\
$B^{0}_{s}$& $5380(108)(^{21}_{18})$
& $5375(103)(^{20}_{21})$\\
$B^{*} - B^{0}$&$32(4)(^{3}_{2})$
&$35(6)(^{3}_{3})$\\
$B^{*}_{s} - B^{0}_{s}$&$29(3)(^{2}_{2})$
&$32(4)(^{3}_{2})$\\
\hline
$\Sigma_{c}$&$2407(32)(^{32}_{37})$&$2452(38)(^{38}_{36})$\\
$\Xi_{c}$&$2440(27)(^{28}_{26})$&$2473(34)(^{34}_{33})$\\
$\Omega_{c}$&$2652(25)(^{27}_{31})$&$2678(33)(^{33}_{31})$\\
$\Sigma^{*}_{c} - \Sigma_{c}$ &$75(20)(^{14}_{12})$
&$86(18)(^{12}_{13})$\\
$\Xi^{*}_{c} - \Xi^{\prime}_{c}$ &$71(18)(^{12}_{9})$
&$81(16)(^{11}_{10})$\\
$\Omega^{*}_{c} - \Omega_{c}$&$65(13)(^{7}_{8})$
&$74(14)(^{8}_{8})$\\
$\Sigma_{c} - \Lambda_{c} $ 
&$128(28)(^{39}_{28})$& $162(36)(^{33}_{26})$\\
$\Xi^{\prime}_{c} - \Xi_{c}$ &$104(19)(^{20}_{23})$&$126(21)(^{15}_{22})$\\
\hline
$\Xi_{cc}$&$3562(47)(^{27}_{25})$&$3588(66)(^{32}_{27})$\\
$\Omega_{cc}$&$3681(44)(^{17}_{19})$&$3698(60)(^{26}_{23})$\\
$\Xi^{*}_{cc} - \Xi_{cc}$ &$63(14)(^{9}_{7})$
&$70(11)(^{7}_{7})$\\
$\Omega^{*}_{cc} - \Omega_{cc}$&$56(8)(^{7}_{6})$
&$63(7)(^{5}_{5})$\\
\hline
$\Lambda_{b}$&$5664(98)(^{33}_{46})$&$5672(102)(^{35}_{41})$\\
$\Xi_{b}$&$5762(83)(^{29}_{38})$&$5788(86)(^{30}_{36})$\\
$\Omega_{b}$&$6021(75)(^{27}_{34})$&$6040(77)(^{25}_{31})$\\
$\Sigma^{*}_{b} - \Sigma_{b}$ &$22(10)(^{7}_{6})$&$24(11)(^{7}_{5})$\\
$\Xi^{*}_{b} - \Xi^{\prime}_{b}$ &$21(10)(^{7}_{6})$&$23(11)(^{7}_{5})$\\
$\Omega^{*}_{b} - \Omega_{b}$&$18(7)(^{6}_{4})$&$20(8)(^{5}_{3})$\\
$\Sigma_{b} - \Lambda_{b}$ &$141(24)(^{30}_{22})$&$175(27)(^{26}_{24})$\\
$\Xi^{\prime}_{b} - \Xi_{b}$ &$124(22)(^{32}_{18})$&$148(25)(^{24}_{15})$\\
\hline
$\Xi^{*}_{bb} - \Xi_{bb}$ &$22(6)(^{4}_{3})$
&$20(6)(^{3}_{4})$\\
$\Omega^{*}_{bb} - \Omega_{bb}$&$20(4)(^{3}_{3})$
&$19(4)(^{3}_{3})$\\
$\Xi^{\prime}_{cb}$ & $6810(150)(^{62}_{79})$
& $6840(228)(^{58}_{72})$\\
$\Omega^{\prime}_{cb}$& $6935(135)(^{75}_{89})$
& $6954(214)(^{62}_{81})$\\
$\Xi^{*}_{cb} - \Xi^{\prime}_{cb}$ &$46(8)(^{4}_{6})$
&$43(9)(^{6}_{6})$\\
$\Omega^{*}_{cb} - \Omega^{\prime}_{cb}$&$40(6)(^{4}_{5})$
&$39(6)(^{5}_{5})$\\
$\Xi_{cb} - \Xi^{\prime}_{cb}$ &$11(6)(^{4}_{5})$&$9(5)(^{6}_{4})$\\
$\Omega_{cb} - \Omega^{\prime}_{cb}$&$10(5)(^{4}_{4})$&$9(4)(^{4}_{4})$\\
\end{tabular}
\end{ruledtabular}
\end{table}

\begin{table}[htb!]
\caption{Results for charmed baryon masses and mass differences
(in MeV) compared to experimental values.
The first row of lattice results (taken from \cite{hv4})
were calculated using a relativistic
action of the D234 type for the charmed quark
while for the second row (this work), the NRQCD action was used.}
\centerline{}
\begin{ruledtabular}
\begin{tabular}{ccccc}
&\multicolumn{3}{c}{Lattice results}&Expt\\
&\( \beta =2.1 \)&
\( \beta =2.3 \)&
\( \beta =2.5 \)&\\
\hline
\( \Sigma _{c} \)&2379(31)\( \left( _{18}^{23}\right)  \)&
2490(14)\( \left( _{33}^{17}\right)  \)&
2493(22)\( \left( _{29}^{21}\right)  \)&
2455\\
&$2407(32)(^{32}_{37})$&$2452(38)(^{38}_{36})$&&\\
\( \Xi _{c} \)&
2455(17)\( \left( _{42}^{11}\right)  \)&
2462(14)\( \left( _{30}^{\, 5}\right)  \)&
2481(14)\( \left( _{34}^{\, 1}\right)  \)&
2468\\
&$2440(27)(^{28}_{26})$&$2473(34)(^{34}_{33})$&&\\
\( \Omega _{c} \)&
2671(11)\( \left( _{59}^{11}\right)  \)&
2699(10)\( \left( _{41}^{\, 8}\right)  \)&
2700(11)\( \left( _{40}^{\, 8}\right)  \)&
2704\\
&$2652(25)(^{27}_{31})$&$2678(33)(^{33}_{31})$&&\\
$\Sigma _{c}^{*}-\Sigma _{c}$&
62(33)\( \left( _{32}^{19}\right)  \)&
82(12)\( \left( _{6}^{9}\right)  \)&
76(19)\( \left( _{\, 4}^{15}\right)  \)&
64\\
&$75(20)(^{14}_{12})$&$86(18)(^{12}_{13})$&&\\
$\Xi _{c}^{*}-\Xi _{c}'$&
52(15)\( \left( _{4}^{8}\right)  \)&
82(10)\( \left( _{5}^{8}\right)  \)&
77(9)\( \left( _{5}^{7}\right)  \)&
70\\
&$71(18)(^{12}_{9})$&$81(16)(^{11}_{10})$&&\\
$\Omega _{c}^{*}-\Omega _{c}$&
50(17)\( \left( _{\, 6}^{11}\right)  \)&
73(8)\( \left( _{5}^{7}\right)  \)&
69(7)\( \left( _{6}^{5}\right)  \)&\\
&$65(13)(^{7}_{8})$&$74(14)(^{8}_{8})$&&\\
\( \Xi _{cc} \)&
3608(15)\( \left( _{35}^{13}\right)  \)&
3595(12)\( \left( _{22}^{21}\right)  \)&
3605(12)\( \left( _{19}^{23}\right)  \)&\\
&$3562(47)(^{27}_{25})$&$3588(66)(^{32}_{27})$&&\\
\( \Omega _{cc} \)&
3747(9)\( \left( _{47}^{11}\right)  \)&
3727(9)\( \left( _{40}^{16}\right)  \)&
3733(9)\( \left( _{38}^{\, 7}\right)  \)&\\
&$3681(44)(^{17}_{19})$&$3698(60)(^{26}_{23})$&&\\
$\Xi _{cc}^{*}-\Xi _{cc}$&
58(14)\( \left( _{10}^{16}\right)  \)&
83(8)\( \left( _{10}^{\, 7}\right)  \)&
80(10)\( \left( _{7}^{3}\right)  \)&\\
&$63(14)(^{9}_{7})$&$70(11)(^{7}_{7})$&&\\
$\Omega _{cc}^{*}-\Omega _{cc}$&
57(8)\( \left( _{\, 9}^{10}\right)  \)&
72(5)\( \left( _{5}^{4}\right)  \)&
68(5)\( \left( _{5}^{6}\right)  \)&\\
&$56(8)(^{7}_{6})$&$63(7)(^{5}_{5})$&&\\
\end{tabular}
\end{ruledtabular}
\end{table}

Moreover, mass differences (Eq.~10) can be calculated more precisely than
masses (Eq.~9).
Therefore, for example, one can calculate a meson mass from the relation
\begin{eqnarray}
M(q_{l},Q)= M(q_h,H) - \Delta M =  M(q_h,H) - \Delta E, 
\end{eqnarray}
where 
\begin{eqnarray}
\Delta M = \Delta E = E(q_h,Q) - E(q_{l},Q).
\end{eqnarray}
Here $q_{h}$ and $q_{l}$ denote the heaviest light quark and a lighter one
respectively, and $M(q_h,H)$ is extracted by using Eq.~(9).
Eq.~(13) is valid as long as $Z$ in Eq.~(11) is the same {\it i.e.,} both
states consist of the same heavy quark $Q$.

Similarly, masses of singly and doubly heavy baryons can be extracted from
meson masses by using
\begin{eqnarray}
M(q_{1}q_{2},Q)     &=& M(q_{h},Q) - \Delta E_{sh},\\
M(q_{1},QQ)     &=& M(QQ) - \Delta E_{dh},
\end{eqnarray}
where
\begin{eqnarray}
E_{sh}   &=& E(q_{h},Q) - E(q_{1}q_{2},Q),\\
E_{dh}    &=&  E(QQ) - E(q_{1},QQ).
\end{eqnarray}
For example, the $\Sigma_{c(b)}$ mass is extracted by taking its difference
(at each $\kappa$) with the $D(B^{0})$ mass ($m$) at $\kappa = 0.229$ and then
subtracting that from $m$. Masses extracted by using Eq.~(9) and Eqs.~(12-17)
are consistent with each other. However, errors in the second method are
smaller than the previous one. 
\begin{figure}[htb!]
\vspace*{-1.35in}
\hspace*{-0.75in}
\includegraphics[height=15.5cm]{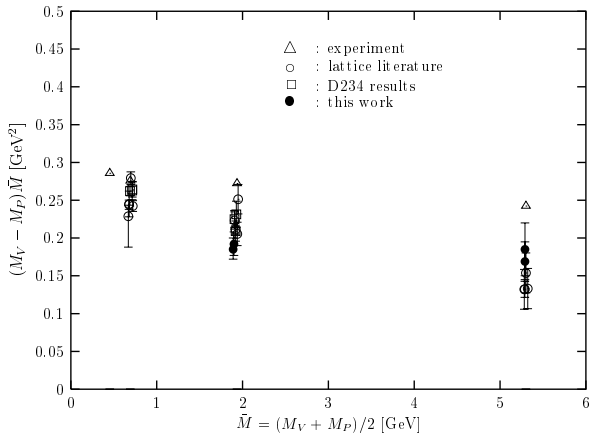}
\vspace*{-2.2in}
\caption{Spin splittings in the meson sector, plotted as 
$(M_V-M_P) \bar M$
versus $\bar M$ where $\bar M$ is the average vector and pseudoscalar
meson mass $(M_V+M_P)/2$.
``Lattice literature'' 
results are from Refs.~\protect\cite{spwave,hypmb,hypmeson} while D234 results
are from Ref~\cite{hv4}. }
\end{figure}

\section{RESULTS}
The mass spectrum and spin splittings of heavy quark baryons have been
computed on an anisotropic lattice using the NRQCD heavy quark action. 
Results are summarized in Table V, where the first error is the statistical
error obtained from a bootstrap analysis with a bootstrap sample size equal
to the configuration sample size. The second error is 
an overall systematic error due to scale and anisotropy uncertainties,
the uncertainty due to 
choosing a time window, the light quark extrapolation 
error and the strange quark mass uncertainty. Mesons, singly heavy baryons 
and doubly heavy baryons are separated into different 
groups by horizontal lines.
In Table VI we have compared our results with those obtained 
by using a relativistic (D234) heavy quark action \cite{hv4}
and experimental numbers (where available).
One can notice that the NRQCD results and D234 results are consistent with 
each other.
Results are also consistent with a previous NRQCD calculation \cite{hv2}.
As in Ref.~\cite{hv4}, it is found that the suppression of 
spin splittings is not present in the baryon sector, although such a suppression
is known to be characteristic of the heavy meson sector.
One can also notice that the spin splittings for 
doubly heavy baryons are as large as their singly heavy counterparts.

\section{DISCUSSION and SUMMARY}
In order to put the results of the present calculation into perspective it
is useful to consider spin splittings over the whole range of available
quark masses. We start with mesons where it has been known for a long
time that the squared mass difference \( M_{V}^{2}-M_{P}^{2} \)
for vector and pseudoscalar mesons is approximately constant for all
mesons of the form \( Q\overline{q} \), where \( q \) is up or down and 
\( Q \) is any light or heavy flavor. This relation is illustrated in
Fig.~4 for the mass pairs \( (\rho ,\pi ), \) \( (K^{*},K), \)
\( (D^{*},D) \) and \( (B^{*},B). \) Also shown are the results 
of lattice simulations including the present work. The tendency for
quenched lattice  QCD to underestimate the spin splittings relative to
experimental values is clearly visible.
\begin{figure}[htb!]
\vspace*{-1.35in}
\hspace*{-0.75in}
\includegraphics[height=15.5cm]{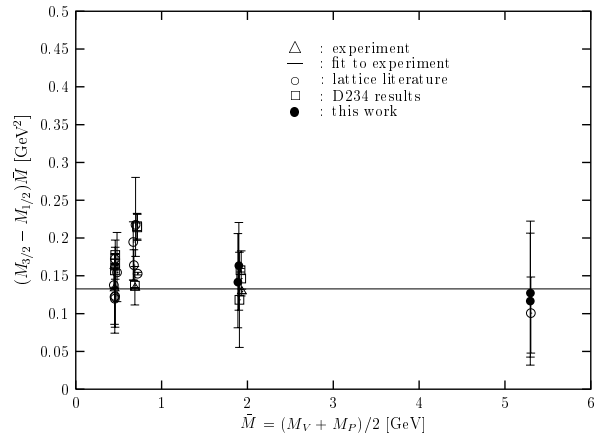}
\vspace*{-2.2in}
\caption{Spin splittings in the baryon sector
plotted as $(M_{3/2}-M_{1/2}) \bar M$
versus $\bar M$ where $\bar M$ is the average vector and pseudoscalar
meson mass $(M_V+M_P)/2$. 
``Lattice literature''
results are from Ref.~\protect\cite{hypmb}
while D234 results are from Ref~\cite{hv4}.
The solid line is a fit to the experimental data}
\end{figure}

In Ref.~\cite{hv4} we showed that it is useful to consider 
the behavior of the 
spin splittings in the baryon sector as a function of quark mass
also in terms of the mesonic average mass \( (M_{V}+M_{P})/2 \).
The results for the baryon pairs
\( (\Delta ,N) \), \( (\Sigma ^{*},\Sigma ) \),
\( (\Sigma _{c}^{*},\Sigma _{c}) \)
and \( (\Sigma _{b}^{*},\Sigma _{b}) \) are shown in Fig.~5.
It is a remarkable empirical fact that the baryon spin splitting scales
almost exactly like the inverse of the average meson mass. The implication 
is that the
ratio of meson to baryon spin splittings is almost constant. This was
discussed in Ref.~\cite{hv4} and was to some extent anticipated by 
Lipkin \cite{Lipkin} from the point of view of the quark model (see also
Lipkin and O'Donnell \cite{LipO}).

The results of quenched lattice calculations are also shown in Fig.~5.
The suppression of spin splittings relative to experiment,
visible for mesons, is not seen
for baryons. The results of the present lattice NRQCD calculation
support this conclusion in the charm and bottom sectors. It is clear that
a definitive measurement of \( \Sigma _{b} \) and \( \Sigma _{b}^{*} \)
masses would be highly desirable to extend the experimental comparison
to larger mass values.

{}From the point of view of lattice NRQCD our results present an interesting
challenge. As is well known, the spin splittings of both 
charmonium \cite{trot}
and heavy-light mesons \cite{quen_res,spwave} 
are clearly underestimated by lattice NRQCD
simulations. Up to now, these simulations have used couplings modified
only by mean-field tadpole factors. It is tempting to speculate that
there are additional large perturbative corrections to these couplings.
In particular it might seem that a correction to the quark coupling
with the chromomagnetic field (the $c_4$-term in Eq.~(A5)) has the 
potential to cure the spin splitting deficiency for both charmonium
and heavy-light mesons. However one has to be cautious in wishing for
such a cure as it would upset the already reasonable values for spin
splittings in the baryon sector.

To summarize, we have calculated the masses of baryons containing 
one or two heavy quarks using quenched lattice QCD. NRQCD is used to
describe charm and bottom quarks. In the charm sector the results
of this work are compatible with those obtained previously where a
Dirac-Wilson action of the D234 type was used for the heavy quark.
No suppression of the spin splittings observed in lattice NRQCD 
simulations of heavy-light mesons is seen in the heavy baryon sector.

This and our previous work \cite{hv4} leave a number of difficult open
questions. One would like to be able to improve the lattice calculations
of baryons to reduce the uncertainties to the same level achievable in mesons.
Also how (and whether) the addition of dynamical quarks to the simulations 
will solve the dilemma of spin splittings has yet
to be understood. A phenomenological
issue is to understand the remarkable constancy in the meson to baryon 
spin splitting ratio over the whole available quark mass range. On the
experimental side it will be a significant challenge to extend
baryon mass measurements in the bottom and doubly heavy sectors.

\begin{acknowledgments}
This work was supported in part by the Natural Sciences and Engineering 
Research Council of Canada. NM is also thankful to CCS of University of Kentucky
for its support.
Some of the computing was done on hardware funded by the Canada Foundation
for Innovation, with contributions from Compaq Canada, Avnet Enterprise
Solutions, and the Government of Saskatchewan.
\end{acknowledgments}

\appendix

\section{Details of Actions}
\subsection{NRQCD Action}
The heavy quark action is nonrelativistic and is discretized to give
the following Green's function propagation:
\begin{eqnarray}
G_{\tau+1} &=& \left(1-\frac{a_tH_B}{2}\right)
    \left(1-\frac{a_tH_A}{2n}\right)^n\nonumber\\
&&\frac{U_4^\dagger}{u_t}
    \left(1-\frac{a_tH_A}{2n}\right)^n\left(1-\frac{a_tH_B}{2}\right) G_\tau,
\end{eqnarray}
The nonrelativistic Hamiltonian
is complete to $O(1/M^3)$ in the classical continuum limit :
\begin{eqnarray}                                  
H &=& H_0 + \delta{H}, \label{H} \\
H_0 &=& \frac{-\Delta^{(2)}}{2M}, \\
\delta{H} &=& \delta{H}^{(1)} + \delta{H}^{(2)} + \delta{H}^{(3)} + O(1/M^4) \\
\delta{H}^{(1)} &=& -\frac{c_4}{u_s^4}\frac{g}{2M}\mbox{{\boldmath$\sigma$}}
                    \cdot\tilde{\bf B} + c_5\frac{a_s^2\Delta^{(4)}}{24M}, \\
\delta{H}^{(2)} &=& \frac{c_2}{u_s^2u_t^2}\frac{ig}{8M^2}(\tilde{\bf \Delta}
         \cdot\tilde{\bf E}-\tilde{\bf E}\cdot\tilde{\bf \Delta})\nonumber\\
&&-\frac{c_3}{u_s^2u_t^2}\frac{g}{8M^2}\mbox{{\boldmath$\sigma$}}
\cdot(\tilde{\bf \Delta}\times\tilde{\bf E}-\tilde{\bf E}\times
           \tilde{\bf \Delta})\nonumber\\
&&\hspace*{1in}             - c_6\frac{a_s(\Delta^{(2)})^2}{16n{\xi}M^2}, \\
\delta{H}^{(3)} &=& -c_1\frac{(\Delta^{(2)})^2}{8M^3}
                    -\frac{c_7}{u_s^4}\frac{g}{8M^3}\left\{\tilde\Delta^{(2)},
           \mbox{{\boldmath$\sigma$}}\cdot\tilde{\bf B}\right\}\nonumber\\
&& -\frac{c_9ig^2}{8M^3}\mbox{{\boldmath$\sigma$}}\cdot
                     \left(\frac{\tilde{\bf E}\times\tilde{\bf E}}{u_s^4u_t^4}
                    +\frac{\tilde{\bf B}\times\tilde{\bf B}}{u_s^8}\right)
                     \nonumber \\
             &&\hspace*{-0.4in} -\frac{c_{10}g^2}{8M^3}\left(\frac{\tilde{\bf E}^2}{u_s^4u_t^4}
                +\frac{\tilde{\bf B}^2}{u_s^8}\right)
                -c_{11}\frac{a_s^2(\Delta^{(2)})^3}{192n^2{\xi^2}M^3}.
\end{eqnarray}
Here a tilde signifies discretization errors have been removed.  
In particular,
\begin{eqnarray}
   \tilde{E}_i &=& \tilde{F}_{4i}, \\
   \tilde{B}_i &=& \frac{1}{2}\epsilon_{ijk}\tilde{F}_{jk},\\
   \tilde{F}_{\mu\nu}(x) &=& \frac{5}{6}F_{\mu\nu}(x)
           - \frac{1}{6u_\mu^2}U_\mu(x)F_{\mu\nu}(x+\hat\mu)U_\mu^\dagger(x)\nonumber \\
&&{\hspace*{-0.86in}}- \frac{1}{6u_\mu^2}U_\mu^\dagger(x-\hat\mu)F_{\mu\nu}(x-\hat\mu)
             U_\mu(x-\hat\mu) - (\mu\leftrightarrow\nu).
\end{eqnarray}

The various spatial lattice derivatives are defined as follows:
\begin{eqnarray}
   a_s\Delta_iG(x) &=& \frac{1}{2u_s}[U_i(x)G(x+\hat\imath)
\nonumber\\
&&\hspace*{0.4in} -U^\dagger_i(x-\hat\imath)G(x-\hat\imath)], \\
   a_s\Delta^{(+)}_iG(x) &=& \frac{U_i(x)}{u_s}G(x+\hat\imath) - G(x), \\
   a_s\Delta^{(-)}_iG(x) &=& G(x) -
              \frac{U^\dagger_i(x-\hat\imath)}{u_s}G(x-\hat\imath), \\
   a_s^2\Delta^{(2)}_iG(x) &=& \frac{U_i(x)}{u_s}G(x+\hat\imath) - 2G(x)
\nonumber\\
&&\hspace*{0.4in}+\frac{U^\dagger_i(x-\hat\imath)}{u_s}G(x-\hat\imath), \\
   \tilde\Delta_i &=& \Delta_i
                   - {a_s^2\over 6} \Delta^{(+)}_i\Delta_i\Delta^{(-)}_i, \\
   \Delta^{(2)} &=& \sum_i \Delta^{(2)}_i \label{Laplacian}, \\
   \tilde \Delta^{(2)} &=& \Delta^{(2)} - {a_s^2 \over 12} \Delta^{(4)}, \\
   \Delta^{(4)} &=& \sum_i \left( \Delta^{(2)}_i \right)^2.
\end{eqnarray}

\subsection{Gauge Field Action}
The leading classical errors of the gauge field action  are quartic in 
lattice spacing.  The action is
\begin{eqnarray}
S_G(U) &=& \frac{5\beta}{3}{\Biggl[} 
          \frac{1}{u_s^4\xi}
             \sum_{\rm ps}\left(1-\frac{1}{3}{\rm ReTr}U_{\rm ps}\right)
\nonumber\\
&&\hspace{0.4in}        - \frac{1}{20u_s^6\xi}
             \sum_{\rm rs}\left(1-\frac{1}{3}{\rm ReTr}U_{\rm rs}\right)
\nonumber\\
     && \hspace{0.1in} + \frac{\xi}{u_s^2u_t^2}
             \sum_{\rm pt}\left(1-\frac{1}{3}{\rm ReTr}U_{\rm pt}\right)
\nonumber\\
&&\hspace{0.1in} - \frac{\xi}{20u_s^4u_t^2}
             \sum_{\rm rst}\left(1-\frac{1}{3}{\rm ReTr}U_{\rm rst}\right)     
\nonumber\\
&&\hspace{-0.1in} - \frac{\xi}{20u_s^2u_t^4}
             \sum_{\rm rts}\left(1-\frac{1}{3}{\rm ReTr}U_{\rm rts}\right)
          {\Biggr]},
\end{eqnarray}
where anisotropic ratio $\xi \equiv a_s/a_t$ and $\beta$ is the lattice
gauge field coupling constant.\\  
{\it ps} : spatial plaquettes\\
{\it rs} : spatial planar 1$\times$2 rectangles,\\
{\it pt} : plaquettes in the temporal-spatial plane,\\
{\it rst(rts)} : rectangles with the long side in a spatial(temporal)
direction.
\subsection{Light Quark Action}
For light quarks, we used a D234 action \cite{alford,hv4} with parameters set
to their tadpole-improved classical values.  
Its leading classical errors are cubic in lattice spacing and the action can be written as
\begin{eqnarray}
S_F(\bar{q},q;U)
    &=& \frac{4\kappa}{3}\sum_{x,i}\left[\frac{1}{u_s\xi^2}D_{1i}(x)
        -\frac{1}{8u_s^2\xi^2}D_{2i}(x)\right]\nonumber\\
&&      + \frac{4\kappa}{3}\sum_x\left[\frac{1}{u_t}D_{1t}(x)
        -\frac{1}{8u_t^2}D_{2t}(x)\right] \nonumber\\
  &&  + \frac{2\kappa}{3u_s^4\xi^2}\sum_{x,i<j}\bar\psi(x)\sigma_{ij}F_{ij}(x)
        \psi(x)\nonumber\\
  &&{\hspace*{-1.0in}}      + \frac{2\kappa}{3u_s^2u_t^2\xi}\sum_{x,i}\bar\psi(x)\sigma_{0i}F_{0i}(x)\psi(x) - \sum_x\bar\psi(x)\psi(x),
\end{eqnarray}
where
\begin{eqnarray}
D_{1i}(x) &=& \bar\psi(x)(1-\xi\gamma_i)U_i(x)\psi(x+\hat{i})\nonumber\\
&&  \hspace*{0.3in}+\bar\psi(x+\hat{i})(1+\xi\gamma_i)U_i^\dagger(x)\psi(x), \\
D_{1t}(x) &=& \bar\psi(x)(1-\gamma_4)U_4(x)\psi(x+\hat{t})\nonumber\\
&&  \hspace*{0.3in}+\bar\psi(x+\hat{t})(1+\gamma_4)U_4^\dagger(x)\psi(x), \\
D_{2i}(x) &=& \bar\psi(x)(1-\xi\gamma_i)U_i(x)U_i(x+\hat{i})\psi(x+2\hat{i})
             \nonumber \\
&&\hspace*{-0.4in}+\, \bar\psi(x+2\hat{i})(1+\xi\gamma_i)U_i^\dagger(x+\hat{i})
              U_i^\dagger(x)\psi(x), \\
D_{2t}(x) &=& \bar\psi(x)(1-\gamma_4)U_4(x)U_4(x+\hat{t})\psi(x+2\hat{t})
             \nonumber \\
&&\hspace*{-0.4in}+\,\bar\psi(x+2\hat{t})(1+\gamma_4)U_4^\dagger(x+\hat{t})
              U_4^\dagger(x)\psi(x), \\
gF_{\mu\nu}(x) &=& \frac{1}{2i}\left(\Omega_{\mu\nu}(x)-\Omega^\dagger_{\mu\nu}
                   (x)\right) \nonumber \\
&&\hspace*{0.2in}- \frac{1}{3}{\rm Im}\left({\rm Tr}\Omega_{\mu
                   \nu}(x)\right), \\
\Omega_{\mu\nu} &=& \frac{-1}{4}
\left[U_\mu(x)U_\nu(x+\hat\mu)U_\mu^\dagger(x+\hat\nu)U_\nu^\dagger(x) \right.
       \nonumber \\
&& \hspace*{-0.3in}~~+U_\nu(x)U_\mu^\dagger(x-\hat\mu+\hat\nu)U_\nu^\dagger(x-\hat\mu)
       U_\mu(x-\hat\mu) \nonumber \\
&&\hspace*{-0.8in} ~~+U_\mu^\dagger(x-\hat\mu)U_\nu^\dagger(x-\hat\mu-\hat\nu)
       U_\mu(x-\hat\mu-\hat\nu)U_\nu(x-\hat\nu) \nonumber \\
&&\hspace*{-0.8in} ~~\left.
       +U_\nu^\dagger(x-\hat\nu)U_\mu(x-\hat\nu)U_\nu(x+\hat\mu-\hat\nu)                                                                                                           U_\mu^\dagger(x) \right].
\end{eqnarray}


\end{document}